\documentclass[11pt,a4paper]{article}
\pdfoutput=1

\usepackage{xparse}
\usepackage{jheppub}
\usepackage{latexsym}
\usepackage{multirow}
\usepackage{color}
\usepackage[usenames,dvipsnames,table]{xcolor}
\usepackage{graphicx}
\usepackage{epsfig}  
\usepackage{epsf}    
\usepackage{dcolumn}
\usepackage{bm}
\usepackage{dcolumn}
\usepackage{textcomp}
\usepackage{float}
\usepackage{subfig}
\usepackage{hypcap}
\usepackage[]{hyperref}
\usepackage{makecell}
\usepackage{color}
\usepackage{pifont}
\usepackage{appendix}
\usepackage{url}

\hypersetup{
  bookmarks=true,         
  unicode=false,          
  pdftoolbar=true,        
 pdfmenubar=true,        
 pdffitwindow=true,     
 pdfstartview={FitH},    
 pdfsubject={Dark Matter},   
 pdfnewwindow=true,      
 pdfcreator={RevTeX},
 colorlinks=true,       
 linkcolor=red,          
 citecolor=blue,        
 filecolor=black,      
 urlcolor=blue,           
  }

\title{Unitarity limits on thermal dark matter in (non-)standard cosmologies} 

\author{Disha Bhatia,} 
\author{Satyanarayan Mukhopadhyay}

\affiliation{School of Physical Sciences, Indian Association for the Cultivation of Science, 2A and 2B Raja S.C. Mullick Road, Kolkata 700 032}
\emailAdd{tpdb@iacs.res.in, tpsnm@iacs.res.in}

\abstract
{Using the upper bound on the inelastic reaction cross-section implied by S-matrix unitarity, we derive the thermally averaged maximum dark matter (DM) annihilation rate for general $k \rightarrow 2$ number-changing reactions, with $k \geq 2$, taking place either entirely within the dark sector, or involving standard model fields. This translates to a maximum mass of the particle saturating the observed DM abundance, which, for dominantly $s$-wave annihilations, is obtained to be around $130$ TeV, $1$ GeV, $7$ MeV and $110$ keV, for $k=2,3,4$ and $5$, respectively, in a radiation dominated Universe, for a real or complex scalar DM stabilized by a minimal symmetry. For modified thermal histories in the pre-big bang nucleosynthesis era, with an intermediate period of matter domination, values of reheating temperature higher than $\mathcal{O}(200)$ GeV  for $k \geq 4$, $\mathcal{O}(1)$ TeV for $k=3$ and $\mathcal{O}(50)$ TeV for $k=2$ are strongly disfavoured by the combined requirements of unitarity and DM relic abundance, for DM freeze-out before reheating.}

\begin{document}
\maketitle
\flushbottom
\section{Introduction}
Model independent bounds on the mass of dark matter (DM) candidates are extremely weak~\cite{Gorbunov:2011zz}. Since DM particles must be confined to galaxies, for bosons their de Broglie wavelength should be smaller than the typical size of the DM-rich dwarf galaxies, which is about 1 kilo parsec.  This leads to a lower bound of around $10^{-22}$ eV on the mass of bosonic DM~\cite{Hu:2000ke, Hui:2016ltb}. For fermionic DM the lower bound is much stronger, about $1$ keV, as the Pauli principle sets a maximum value of the phase-space density~\cite{Tremaine:1979we}. The stability of stellar clusters in galaxies also requires a very large upper bound of $10^3$ solar masses on the mass of any DM candidate~\cite{Moore:1993sv, Carr:1997cn}. The allowed range of DM mass may be further restricted only if specific properties of DM particles are assumed. One such well-known example is the assumption that DM was in kinetic equilibrium in the early Universe with the standard model (SM) thermal bath. The free-streaming of such thermal dark matter is constrained by the Lyman-$\alpha$ flux-power spectra data, resulting in a lower limit of around $5.3$ keV~\cite{Irsic:2017ixq}. 

Specifying a particular production mechanism of the DM abundance in the early Universe may indicate a viable DM mass range for that scenario, though the allowed values tend to be model dependent. For DM particles that were in kinetic and chemical equilibrium in the early Universe with the SM bath, and whose present mass density was determined by number-changing pair annihilations to the SM sector, a {\em model independent} upper bound on the DM mass can be obtained from the requirements of unitarity of the S-matrix~\cite{Griest:1989wd, Hui:2001wy}. As first shown by Griest and Kamionkowski~\cite{Griest:1989wd}, such a bound follows from the maximum inelastic reaction rate implied by unitarity, which sets the minimum frozen out number density of DM, and hence the maximum mass that saturates the present density. The presence of long-range interactions which may lead to bound-state formation reduces the effective number-changing inelastic annihilation rate, and thus reduces the unitarity upper limit compared to the scenario with no bound-state effects~\cite{vonHarling:2014kha, Baldes:2017gzw, Smirnov:2019ngs}. Furthermore, the presence of a particle anti-particle asymmetry in the DM sector necessarily implies a non-zero equilibrium chemical potential for such DM, thereby increasing the effective number density of the surviving species at freeze-out, and reducing the unitarity upper limits further~\cite{Baldes:2017gzw, Ghosh:2020lma}. In particular, as shown in Ref.~\cite{Ghosh:2020lma}, in a purely asymmetric DM scenario, generated by a semi-annihilation process with large CP-violation, the unitarity limit on the DM mass can be as strong as $15$ GeV. 

The above studies on unitarity limits have focussed on scenarios in which the dominant DM number changing reaction is of $2 \rightarrow 2$ type, in which a pair of DM particles annihilate either to a pair of SM particles (pair-annihilation), or to a DM and a SM particle (semi-annihilation). However, as pointed out by Carlson, Machacek and Hall~\cite{Carlson:1992fn}, and subsequently revived in recent studies~\cite{Hochberg:2014dra}, the dominant number changing interactions may take place entirely within the dark sector as well. The minimal such reaction involving a single DM species is of $3 \rightarrow 2$ type, dubbed as strongly interacting massive particles (SIMP), due to their appearance in theories of the dark sector involving new strong interactions~\cite{Hochberg:2014kqa}. Generically, a simple low energy effective theory of a complex scalar particle with a cubic and quartic self-interaction will have $3 \rightarrow 2$ number changing reactions.

A natural question therefore is what are the implications of S-matrix unitarity in such scenarios, where the dominant DM number changing interaction is of the type $k \rightarrow n $, with $k \geq n$? A completely general formulation of this problem is challenging since the partial-wave decomposition of a $k-$body initial state, for $k \geq 3$, is rather involved. One might however restrict to a smaller subset of reactions of the $k \rightarrow 2$ type, with $k \geq 2$, since as we shall see, in a thermal bath the relevant thermally averaged reaction rates $\langle \sigma v_{\rm rel}^{k-1} \rangle_{k\rightarrow 2}$ get related to $\langle \sigma v_{\rm rel} \rangle_{2\rightarrow k}$ in equilibrium, and one can of course easily perform a partial wave analysis for the two-body initial state. We therefore only need to find out the maximum value of the inelastic $2 \rightarrow k$ cross-sections as implied by unitarity, which can be obtained using the optical theorem and the study of the corresponding $2 \rightarrow 2$ elastic scattering process~\cite{Weinberg:1995mt}. 

A general expression for the thermally averaged maximum rate of $k \rightarrow 2$ reactions $\langle \sigma v_{\rm rel}^{k-1} \rangle_{k\rightarrow 2}$ is the primary result obtained in this paper, as detailed in Secs.~\ref{sec_unitarity} and ~\ref{sec_Boltzmann}. The modification to the upper bound for two identical initial state particles in the $2 \rightarrow k$ reaction is also given. Having obtained the upper limit on the annihilation rate, we translate this to a bound on the maximum mass of the DM particle in a radiation dominated Universe in Sec.~\ref{sec_limits_RD}, using the Boltzmann equation for $k \rightarrow 2$ reactions set up in Sec.~\ref{sec_Boltzmann}. 
Sec.~\ref{sec_limits_RD} provides the generalization of the bound of Ref.~\cite{Griest:1989wd} for $k \rightarrow 2$ reactions taking place within the dark sector. We then go on to consider the possibility of a modification to the thermal history of the Universe, taking up the frequently occuring scenario of an intermediate period of matter domination in the pre-big bang nucleosynthesis (BBN) era~\cite{Kolb:1990vq, Kamionkowski:1990ni, McDonald:1989jd, Gelmini:2006pw,  Allahverdi:2020bys, Drees:2017iod, Evans:2019jcs, Arias:2019uol, Hamdan:2017psw}. It is observed in Sec.~\ref{sec_limits_IMD} that though the unitarity limits are weaker in such scenarios primarily due to the dilution of DM density from late-time entropy production, there are strong implications of unitarity to the possible values of the reheating temperature at which the radiation dominated Universe is restored. We summarize our results in Sec.~\ref{sec_summary}.

\section{Implications of S-matrix unitarity}
\label{sec_unitarity}
To begin with, we recall some of the basic results on the implications of S-matrix unitarity, following the treatment of Weinberg in Ref.~\cite{Weinberg:1995mt}. We focus the discussion to the context of dark matter annihilations, and adopt a multi-particle momentum eigenstate  normalization convention different from Ref.~\cite{Weinberg:1995mt}~\footnote{In our convention, the two-particle spin-0 states are normalized as 
\begin{equation}
\begin{split}
\langle \vec{p_1^\prime}, n_1^\prime; \vec{p_2^\prime}, n_2^\prime |  \vec{p_1},n_1; \vec{p_2},n_2  \rangle &= \left(2\pi \right)^3 2E_{\vec{p_1}} \left(2\pi \right)^3 2E_{\vec{p_2}} \delta^{(3)} \left(\vec{p_1}-\vec{p_1^\prime}  \right) \delta^{(3)} \left(\vec{p_2}-\vec{p_2^\prime}  \right) \delta_{n_1^\prime,n_1} \delta_{n_2^\prime, n_2} \\
& + {\rm permutation, ~where ~n_1, n_1^\prime, n_2, n_2^\prime ~are ~the ~particle ~labels}. \nonumber
\end{split}
\end{equation}
}. A similar approach is also followed by Hui~\cite{Hui:2001wy}. 

For a two-particle initial state $\alpha$, unitarity of the S-matrix implies the optical theorem, which reads in the centre of momentum (CM) frame as~\cite{Weinberg:1995mt}
\begin{equation}
\operatorname{Im}{\mathcal{M}_{\alpha \alpha}} = 2 |\vec{p}| E_{\rm CM} \sigma_{\rm tot},
\label{Eq_optical}
\end{equation}
where, $|\vec{p}|$ is the magnitude of the three momentum of each initial particle and $E_{\rm CM}$ is the total initial state energy, both  in the CM frame. The state label $\alpha$ is described by the three momenta, spin-$z$ components (or helicity) of each particle, and all other internal quantum numbers that label the particles, and $\sigma_{\rm tot} = \sum_{\beta}\sigma_{\alpha \rightarrow \beta}$, where $\beta$ denotes all possible final states that can be obtained from the initial state $\alpha$. 

Let us consider for simplicity collision of two non-identical spin$-0$ particles; the generalization to non-zero spin is straightforward~\cite{Weinberg:1995mt, Hui:2001wy}~\footnote{For treatments using the helicity basis, see, Refs.~\cite{Jacob:1959at, martin}.}. We shall discuss the case for identical particles subsequently. To utilize the rotational invariance of the problem, we perform the basis transformation from the momentum eigenstates $|\vec{p_1}, \vec{p_2}, n \rangle$ to the states $|\vec{P}, E, \ell, m, n \rangle$, where $\vec{P}=\vec{p_1}+\vec{p_2}$, $E=E_1+E_2$, and $(\ell,m)$ are a pair of integers such that
\begin{equation}
\langle \vec{p_1}, \vec{p_2}, n^\prime | \vec{P}, E, \ell, m, n \rangle =16 \pi^3 \sqrt{\frac{E}{|\vec{p_1}|}} \delta^{(3)} \left(\vec{P}-\vec{p_1}-\vec{p_2}  \right) \delta \left(E-E_1-E_2  \right) Y_l^m(\hat{p_1}) \delta_{n^\prime, n},
\label{Eq_norm_1}
\end{equation}
where $Y_l^m(\hat{p_1})$ are the spherical harmonics corresponding to the direction unit vector $\hat{p_1}$. Here $\ell$ corresponds to the total orbital angular momentum of the two initial state particles in the CM frame, and $m$ to its $z-$component. The channel index $n$ stands for the two particle species' labels $n_1$ and $n_2$, and similarly for $n^\prime$. The normalization of the scalar product is chosen to ensure that in the CM frame, the state vector $|0, E, \ell, m, n \rangle$ has the following inner product with a general state vector $|\vec{P^\prime}, E^\prime, \ell^\prime, m^\prime, n^\prime \rangle$
\begin{equation}
\langle \vec{P^\prime}, E^\prime, \ell^\prime, m^\prime, n^\prime | 0, E, \ell, m, n \rangle = \delta^{(3)} \left(\vec{P^\prime}  \right) \delta \left(E^\prime-E  \right) \delta_{\ell,\ell^\prime} \delta_{m,m^\prime} \delta_{n^\prime, n}.
\label{Eq_norm_2}
\end{equation}
With this, we can write the matrix elements for the operator $S-I$ as 
\begin{equation}
\langle \vec{P^\prime}, E^\prime, \ell^\prime, m^\prime, n^\prime | S-I | 0, E, \ell, m, n \rangle = \delta^{(3)} \left(\vec{P^\prime}  \right) \delta \left(E^\prime-E  \right) \delta_{\ell,\ell^\prime} \delta_{m,m^\prime} \left(S_{n^\prime, n}(\ell,E)- \delta_{n^\prime, n}\right),
\end{equation}
where, we have subtracted out the trivial no-scattering part of the S-matrix, and consider only the connected part of the S-matrix operator, $S-I$. The matrix element is independent of $m$ and is only a function of $\ell$ and $E$, as the operator $S$ commutes with the generators of rotation. With $S_{\beta \alpha}=\delta\left(\beta-\alpha\right)+ \left(2\pi\right)^4 \delta^{(4)} \left(p_\beta - p_\alpha\right) (i \mathcal{M}_{\beta \alpha})$, we then obtain for a $2\rightarrow 2$ scattering with $ |  \vec{p_1}, \vec{p_2}, n \rangle \rightarrow |\vec{p_1^\prime}, \vec{p_2^\prime}, n^\prime \rangle$, in the CM frame,
\begin{equation}
\mathcal{M}_{n^\prime, n}= -i \frac{16\pi^2 E}{\sqrt{|\vec{p_1}| |\vec{p_1^\prime}|}}\sum_{\ell,m} Y_l^m(\hat{p_1^\prime}) Y_l^m(\hat{p_1})^* \left(S_{n^\prime, n}(\ell,E)- \delta_{n^\prime, n}\right).
\label{Eq_ME_Gen}
\end{equation}
Choosing $\hat{p_1}$ along the $z-$direction, and integrating over $d\Omega(\hat{p_1^\prime})$, one obtains for the elastic $2\rightarrow 2$  cross-section in the CM frame in the channel $n$
\begin{equation}
\sigma_{\rm elastic} = \sum_{\ell} \frac{\pi}{|\vec{p_1}|^2} \left(2\ell+1\right) |\left(S_{n, n}(\ell,E)-1\right)|^2.
\label{Eq_elastic}
\end{equation}
Now using the optical theorem in Eq.~\ref{Eq_optical} and the general expression for the matrix element in Eq.~\ref{Eq_ME_Gen}, we find the total cross-section from the channel $n$, which includes both elastic and inelastic processes, as
\begin{equation}
\sigma_{\rm total} =\sum_{\ell} \frac{\pi}{|\vec{p_1}|^2} \left(2\ell+1\right) 2\operatorname{Re}\left(1-S_{n, n}(\ell,E)\right).
\label{Eq_sigma_total}
\end{equation}
Finally, subtracting the elastic cross-section in Eq.~\ref{Eq_elastic} from the total cross-section in Eq.~\ref{Eq_sigma_total} gives us the total inelastic scattering cross-section from the channel $n$
\begin{equation}
\sigma_{\rm inelastic} =\sum_{\ell} \frac{\pi}{|\vec{p_1}|^2} \left(2\ell+1\right) \left(1-|S_{n, n}(\ell,E)|^2 \right).
\label{Eq_sigma_inelastic}
\end{equation}
As $|S_{n, n}(\ell,E)|^2 \geq 0$, we obtain the upper bound on the total inelastic cross-section implied by unitarity
\begin{equation}
\sigma_{\rm inelastic} \leq \sum_{\ell} \frac{\pi}{|\vec{p_1}|^2} \left(2\ell+1\right),
\label{Eq_inelastic_bound}
\end{equation}
as also derived in~\cite{Hui:2001wy}. For the annihilation of a pair of non-relativistic DM particles of mass $m_\chi$ such that $|\vec{p_1}| \simeq m_\chi v_{\rm rel}/2$, with $v_{\rm rel}$ being the relative velocity of the colliding particles, Eq.~\ref{Eq_inelastic_bound} implies the well-known result~\cite{Griest:1989wd}
\begin{equation}
\sigma_{\rm inelastic} \leq \sum_{\ell} \frac{4\pi}{m_\chi^2 v_{\rm rel}^2} \left(2\ell+1\right),
\end{equation}
assuming that the $2 \rightarrow 2$ process under consideration saturates the unitarity upper limit.

So far, the discussion was focussed on non-identical initial state particles. For {\em identical particles in the initial state}, we modify the normalization in Eq.~\ref{Eq_norm_1} with an additional factor of $\sqrt{2}$,
\begin{equation}
\begin{split}
        \langle \vec{p_1}, \vec{p_2}, n^\prime | \vec{P}, E, \ell, m, n \rangle &= 16 \sqrt{2} \pi^3 \sqrt{\frac{E}{|\vec{p_1}|}} \delta^{(3)} \left(\vec{P}-\vec{p_1}-\vec{p_2}  \right) \delta \left(E-E_1-E_2  \right) Y_l^m(\hat{p_1}) \delta_{n^\prime,n} \\
        & {\rm \hspace{5cm}(for ~identical ~initial ~particles}),
\end{split}
\label{Eq_norm_1_identical}
\end{equation}
such that the angular momentum eigenstate normalization remains the same as in Eq.~\ref{Eq_norm_2}. With this the expression for elastic scattering cross-section in the channel $n$ becomes
\begin{equation}
\sigma_{\rm elastic} = \sum_{\ell} \frac{2\pi}{|\vec{p_1}|^2} \left(2\ell+1\right) |\left(S_{n, n}(\ell,E)-1\right)|^2~~{\rm (for ~identical ~initial ~particles}).
\label{Eq_elastic_identical}
\end{equation}
Similarly, the expression for total cross-section from the channel $n$ also gets modified to 
\begin{equation}
\sigma_{\rm total} =\sum_{\ell} \frac{2\pi}{|\vec{p_1}|^2} \left(2\ell+1\right) 2\operatorname{Re}\left(1-S_{n, n}(\ell,E)\right)~~{\rm (for ~identical ~initial ~particles}), 
\label{Eq_sigma_total_identical}
\end{equation}
thus modifying the total inelastic cross-section from the channel $n$ and its corresponding upper bound
\begin{align}
\label{Eq_sigma_inelastic_identical}
\sigma_{\rm inelastic} & = \sum_{\ell} \frac{2\pi}{|\vec{p_1}|^2} \left(2\ell+1\right) \left(1-|S_{n, n}(\ell,E)|^2 \right) \nonumber \\
                                      & \leq \sum_{\ell} \frac{2\pi}{|\vec{p_1}|^2} \left(2\ell+1\right)~~{\rm (for ~identical ~initial ~particles}).
\end{align}
Our results for the maximum value of the inelastic scattering cross-section for identical initial state particles does not agree with the corresponding comment in Ref.~\cite{Hui:2001wy}, but agrees with the appearance of this extra symmetry factor (of 2, for 2 identical particles in the initial state) with Ref.~\cite{Namjoo:2018oyn}. However, the maximum values of the inelastic $k\rightarrow 2$ reaction rates obtained in  Refs.~\cite{Namjoo:2018oyn, Kuflik:2017iqs} do not agree with our results in the next section, as the authors in Refs.~\cite{Namjoo:2018oyn, Kuflik:2017iqs} did not maximize the inelastic rate while taking into account the fact that for a non-zero inelastic reaction rate, the elastic scattering cross-section is always non-zero as well~\cite{Landau:1991wop, Weinberg:1995mt, Hui:2001wy}. As seen in this section, our derivation of the maximum rate of inelastic reaction cross-sections uses only the optical theorem and the matrix element and cross-section for the relevant $2 \rightarrow 2$ elastic scattering process.

\section{General Boltzmann equations for $k \rightarrow 2$ DM annihilation processes}
\label{sec_Boltzmann}
We now consider the implications of the upper bound from unitarity on the total inelastic reaction rate, on the number-changing dark matter annihilations in the early Universe. The evolution of the number density of dark matter particles in the expanding universe is described by a set of coupled Boltzmann equations. With general $k\rightarrow 2$ collision reactions, for $k\geq 2$, the number density of the $i-$th particle $n_i$ satisfies the equation
\begin{equation}
\frac{dn_i}{dt}+3Hn_i = - \sum_{\rm Channels} \Delta n_i \int \prod_{\alpha } d\Pi_\alpha \left(2 \pi\right)^4  \delta^4\left(p_1+p_2+p_3+...+p_k - p_a -p_b\right) \omega\left(\vec{p_\alpha},t \right),
\label{Eq_Boltz_1}
\end{equation}
where $H$ is the Hubble expansion rate of the Universe. The product over $\alpha$ includes the momentum integral factors $d\Pi_\alpha$ for all the $(n+2)$ particles, and the sum over the reaction channels indicates all possible reactions involving the production and destruction of the $i-$th particle, with the net change in the $i-$th particle number being $\Delta n_i $ in a reaction. Appropriate symmetry factors should be included in the momentum integrals to take into account the presence of identical particles. Here, we define
\begin{equation}
d\Pi_\alpha = \frac{d^3 p_\alpha}{\left(2 \pi\right)^3 2 E_{\vec{p_\alpha}}},
\end{equation}
and the number density of each species $n_\alpha(t)$ is given by an integral over the distribution function $f_\alpha({\vec{p_\alpha}},t)$
\begin{equation}
n_\alpha(t) = \frac{g_\alpha}{\left(2 \pi\right)^3} \int f_\alpha(\vec{p_\alpha},t) d^3 p_\alpha.
\end{equation}
The function $\omega\left(\vec{p_\alpha}\right)$, ignoring Pauli blocking and Bose enhancement factors, is given as 
\begin{equation}
\omega\left(\vec{p_\alpha}, t \right) = \prod_{\alpha=1}^k f_\alpha(\vec{p_\alpha},t) \sum_{\rm spins} |\mathcal{M}|^2_{1+2+...+k \rightarrow a+b} - f_a(\vec{p_a},t)  f_b(\vec{p_b},t) \sum_{\rm spins} |\mathcal{M}|^2_{a+b \rightarrow 1+2+...+k},
\end{equation}
where, the matrix elements have been summed over both initial and final spins. We can rewrite Eq.~\ref{Eq_Boltz_1} in terms of the unpolarized cross-sections of the above reactions as
\begin{equation}
\frac{dn_i}{dt}+3Hn_i = - \sum_{\rm Channels} \Delta n_i \left[n_1 n_2 ... n_k \langle \sigma_{k \rightarrow 2} v_{\rm rel}^{k-1} \rangle - n_a n_b \langle \sigma_{2 \rightarrow k} v_{\rm rel} \rangle \right],
\label{Eq_Boltzmann_number}
\end{equation}
where, the thermally averaged reaction rate is given by
\begin{equation}
\langle \sigma_{k \rightarrow 2} v_{\rm rel}^{k-1} \rangle = \frac{\int d^3 p_1 ... d^3p_k f_1^{\rm eq} ... f_k^{\rm eq} \sigma_{k \rightarrow 2} v_{\rm rel}^{k-1}}{\int d^3 p_1 ... d^3p_k f_1^{\rm eq} ... f_k^{\rm eq}},
\label{Eq_thermal_average}
\end{equation}
where, $\sigma_{k \rightarrow 2}$ is the unpolarized (summed over final spins and averaged over initial spins) cross-section for the ${k \rightarrow 2}$ process, $v_{\rm rel}$ is the relative velocity of each particle pair, and $f_\alpha^{\rm eq}$ is the equilibrium distribution function for the particle species $\alpha$. The appropriate symmetry factors appearing in Eq.~\ref{Eq_Boltz_1} should now be included in Eq.~\ref{Eq_thermal_average}. The thermal average $\langle \sigma_{2 \rightarrow k} v_{\rm rel} \rangle$ can be similarly performed. Eq.~\ref{Eq_Boltzmann_number} can be further simplified by noting that in equilibrium the following relation is satisfied for each individual reaction channel:
\begin{equation}
n_1^{\rm eq} n_2^{\rm eq} ... n_k^{\rm eq} \langle \sigma_{k \rightarrow 2} v_{\rm rel}^{k-1} \rangle = n_a^{\rm eq} n_b^{\rm eq} \langle \sigma_{2 \rightarrow k} v_{\rm rel} \rangle.
\label{Eq_detailed_balance}
\end{equation}
Thus, we can express the Boltzmann equations only in terms of $\langle \sigma_{2 \rightarrow k} v_{\rm rel} \rangle$ by using Eq.~\ref{Eq_detailed_balance} in Eq.~\ref{Eq_Boltzmann_number}
\begin{equation}
\frac{dn_i}{dt}+3Hn_i = - \sum_{\rm Channels} (\Delta n_i)  n_a^{\rm eq} n_b^{\rm eq} \langle \sigma_{2 \rightarrow k} v_{\rm rel} \rangle \left[\frac{n_1 n_2 ... n_k}{n_1^{\rm eq} n_2^{\rm eq} ... n_k^{\rm eq} }  -\frac{ n_a n_b}{n_a^{\rm eq} n_b^{\rm eq}}  \right]
\label{Eq_Boltzmann_2tok}
\end{equation}

We can easily perform the thermal average integral in Eq.~\ref{Eq_thermal_average} for the $2\rightarrow k$ reactions with the maximum value of the inelastic cross-section as in Eq.~\ref{Eq_inelastic_bound} as input, and obtain, with all $k+2$ particles having the same mass,
\begin{equation}
\langle \sigma_{2 \rightarrow k} v_{\rm rel} \rangle_{\rm max}=\sum_{\ell} \left(2\ell +1\right)  \frac{4 \sqrt{\pi}}{m_\chi^2} \sqrt{x} e^{-\left(k-2\right)x},
\label{Eq_sigmav_average_max}
\end{equation} 
where, $x=m_\chi/T$. For the DM scenarios to be considered subsequently, the $k+2$ particles will be of the same species, including anti-particles, and will thus have the same mass, $m_\chi$. Here we have assumed that the unitarity limit on the total inelastic cross-section is saturated by the specific $2\rightarrow k$ reaction under consideration. With that, for $k=2$ and $\ell=0$ this  reduces to the well-known result~\cite{Griest:1989wd}
\begin{equation}
\langle \sigma_{2 \rightarrow 2} v_{\rm rel} \rangle_{\rm max, ~s-wave}=  \frac{4 \sqrt{\pi}}{m_\chi^2} \sqrt{x}.
\label{Eq_sigma_max_2to2}
\end{equation} 
For $k \geq 3$ we see that there is an exponential suppression factor in $\langle \sigma_{2 \rightarrow k} v_{\rm rel} \rangle^{\rm max}$, namely $e^{-\left(k-2\right)x}$, due to the phase-space cost for producing each extra particle. We can now use Eq.~\ref{Eq_detailed_balance} to obtain the maximum value of the thermally averaged rate $\langle \sigma_{k \rightarrow 2} v_{\rm rel}^{k-1} \rangle$ as follows:
\begin{equation}
\langle \sigma_{k \rightarrow 2} v_{\rm rel}^{k-1} \rangle_{\rm max}=\sum_{\ell} \left(2\ell +1\right)  \frac{2^{\frac{3k-2}{2}} \left(\pi x \right)^{\frac{3k-5}{2}}}{g_\chi^{k-2} m_\chi^{3k-4}} .
\label{Eq_sigmav_kto2_average_max}
\end{equation}
Here $g_\chi$ is the number of spin degrees of freedom of the DM particle. Thus, for example, the maximum value of the thermally averaged s-wave cross-section for a $3\rightarrow 2$ reaction is given by 
\begin{equation}
\langle \sigma_{3 \rightarrow 2} v_{\rm rel}^{2} \rangle_{\rm max,~s-wave}=  \frac{8\sqrt{2} \left(\pi x \right)^{2}}{g_\chi m_\chi^{5}} .
\label{Eq_3to2_sigma_max}
\end{equation}
Similarly, the maximum value of the thermally averaged s-wave cross-section for a $4\rightarrow 2$ reaction is given by
\begin{equation}
\langle \sigma_{4 \rightarrow 2} v_{\rm rel}^{3} \rangle_{\rm max,~s-wave}=  \frac{32 \left(\pi x \right)^{7/2}}{g_\chi^2 m_\chi^{8}} .
\label{Eq_4to2_sigma_max}
\end{equation}
As mentioned at the end of the previous section, our results for the thermally averaged reaction rates in Eqs.~\ref{Eq_sigmav_kto2_average_max}, ~\ref{Eq_3to2_sigma_max} and ~\ref{Eq_4to2_sigma_max} do not agree with the results in Refs.~\cite{Namjoo:2018oyn, Kuflik:2017iqs}.

For identical particles in the initial state, we found in Eq.~\ref{Eq_sigma_inelastic_identical} that the maximum inelastic cross-section for a $2 \rightarrow k$ reaction is a factor of two larger than the non-identical case. However, the thermal averaging integral in $\langle \sigma_{2 \rightarrow k} v_{\rm rel} \rangle$ will in this case have a symmetry factor of $1/2$ for the two identical particles in the initial state. Therefore, for identical initial state particles, $\langle \sigma_{2 \rightarrow k} v_{\rm rel} \rangle_{\rm max}$ as shown in Eq.~\ref{Eq_sigmav_average_max} remains valid, and similarly Eq.~\ref{Eq_sigmav_kto2_average_max} remains the same as well. 

\section{Unitarity limits on thermal DM mass: radiation dominated Universe}
\label{sec_limits_RD}
We can now apply the results on the unitarity upper bound for the thermally averaged annihilation rates to find out the limits on the DM mass. To begin with, we consider the standard scenario of a radiation dominated Universe in the early epochs before big-bang nucleosynthesis. In such a Universe, the Boltzmann equation for a dark matter particle thermalized with the SM bath, going through $k \rightarrow 2$ annihilations within the same species (i.e., when all the $k+2$ particles involved in the collision are the same), can be written using  Eq.~\ref{Eq_Boltzmann_2tok} as
\begin{equation}
\frac{dY}{dx} =  - \sum_{\rm Channels} (\Delta n_\chi) \frac{s \langle \sigma_{2 \rightarrow k} v_{\rm rel} \rangle}{xH Y_{\rm eq}^{k-2}} \left[Y^k - Y^2 Y_{\rm eq}^{k-2} \right],
\label{Eq_Y_isoentropic}
\end{equation}
where, we have defined $Y = n_\chi/s$ and $x=m_\chi/T$, with $s$ being the entropy density, with the assumption that it is conserved during the evolution considered above, i.e., $s(t) a(t)^3={\rm constant}$, where $a(t)$ is the scale factor describing the expansion of the Universe in the Friedmann-Robertson-Walker cosmology. Taking the maximum value of the reaction rate using Eq.~\ref{Eq_sigmav_average_max}, we then arrive at the equation
\begin{equation}
\frac{dY}{dx} =  - \sum_{\rm Channels} (\Delta n_\chi) \frac{\lambda_R}{x^{3(k-1)/2}} \left[Y^k - Y^2 Y_{\rm eq}^{k-2} \right],
\label{Eq_Y_RD}
\end{equation}
where, $\lambda_R$ is given by
\begin{equation}
\lambda_R = \frac{(2\pi)^{\frac{3(k-2)}{2}}  \sigma_0 s(m_\chi)^{k-1}}{H(m_\chi) g_\chi^{k-2} m^{3(k-2)}} \left(\frac{g_{*,s}(T)}{g_{*,s}(m)}\right)^{k-1} \sqrt{\frac{g_*(m)}{g_*(T)}}.
\label{Eq_lambda_R}
\end{equation}
Here, $\sigma_0=4\sqrt{\pi}/m_\chi^2$, $g_{*,s}(T)$ and $g_*(T)$ are the effective number of relativistic degrees of freedom relevant for the entropy density and Hubble rate, respectively, and $s(m_\chi)$ and $H(m_\chi)$ are the entropy density and Hubble rate evaluated at the temperature $T=m_\chi$.

Eq.~\ref{Eq_Y_RD} can be solved analytically with the approximation that for large values of $x$, which correspond to the late time Universe, we may ignore the $Y_{\rm eq}^{k-2}$ term compared to the $Y^{k-2}$ term, and that the relativistic degrees of freedom do not change appreciably during the DM freeze-out. Then, with the boundary condition $Y(x_F) \simeq Y_{\rm eq}(x_F)$, where, $x_F=m_\chi/T_F$, $T_F$ being the freeze-out temperature, we have 
\begin{equation}
Y_{\rm eq}(x_F) \simeq \frac{x_F H(T_F)}{s(T_F) \langle \sigma_{2 \rightarrow k} v_{\rm rel} \rangle|_{x=x_F}},
\end{equation}
where, $\langle \sigma_{2 \rightarrow k} v_{\rm rel} \rangle$ is to be evaluated at $x=x_F$. With this boundary condition, the approximate solution for the present DM yield is obtained to be
\begin{equation}
Y(x \rightarrow \infty) \simeq \left(\frac{3k-5}{\lambda_R(2k-2)}\right)^{\frac{1}{k-1}} x_F^{\frac{3k-5}{2k-2}}.
\end{equation}
We can thus obtain the approximate value of the DM relic abundance as $\Omega_\chi = m_\chi Y(x \rightarrow \infty) s_0/\rho_c$, where $s_0$ is the present entropy density and $\rho_c$ is the critical density. 

In order to obtain accurate values of the upper bound on the DM mass, we solve the Boltzmann equation~\ref{Eq_Y_isoentropic} numerically in several simple scenarios in which the dominant DM number changing topology is of $k\rightarrow 2$ type, with $k \geq 2$. In each case a minimal scenario with (real or complex) scalar is considered, with a minimal $Z_N$ DM stabilization symmetry. The scenarios are as follows:
\begin{enumerate}
\item $\chi + \overline{\chi} \to \text{SM} + {\text{SM}}$: Here, $\chi$ is a non-self-conjugate DM candidate with a distinct anti-particle $\overline{\chi}$, stabilized by a $Z_2$ symmetry. The Boltzmann equation in this case is the familiar one
\begin{equation}
\frac{1}{a^3} \frac{d }{dt}(n_\chi a^3) =  - \langle \sigma_{\chi + \overline{\chi} \to \text{SM} + {\text{SM}} } v_{\rm rel} \rangle 
\left[ n_\chi^2 - n_{\chi,\text{eq}}^2  \right]\;.
\end{equation}

\item $3 \chi \to 2 \chi$:
Here, $\chi$ is a complex scalar DM stabilized by a $Z_3$ symmetry. Assuming CP-conservation, the Boltzmann equation for the DM number density, taking into account all reactions allowed by the $Z_3$ symmetry, can be written as
\begin{eqnarray}
\frac{1}{a^3} \frac{d }{dt}(n_\chi a^3) &=&  - \dfrac{1}{n_{\chi,\text{eq}}}
 \bigg(  \langle \sigma_{ \chi \chi^* \to \chi \chi \chi  } v \rangle 
+  \langle \sigma_{ \chi^* \chi^* \to \chi \chi \chi^*  } v \rangle \bigg) 
 \times
\left[ n_\chi^3 - n_\chi^2 n_{\chi,\text{eq}}  \right]\;.
\end{eqnarray}

\item $4 \chi \to 2 \chi$: 
Here, $\chi$ is a real scalar DM stabilized by a $Z_2$ symmetry. The Boltzmann equation in this case, taking into account all reactions allowed by the $Z_2$ symmetry, is
\begin{eqnarray}
\frac{1}{a^3} \frac{d }{dt}(n_\chi a^3) =  
- \dfrac{2}{n^2_{\chi,\text{eq}}}  \langle \sigma_{ \chi \chi \to \chi \chi \chi \chi } v \rangle
\bigg[ n_\chi^4 - n_\chi^2 n^2_{\chi,\text{eq}}  \bigg] \;.
\end{eqnarray}

\item $5 \chi \to 2 \chi$ : 
Here, $\chi$ is a complex scalar DM stabilized by a $Z_7$ symmetry (a $Z_5$ symmetry instead would have allowed a $3 \rightarrow 2$ reaction as well). Assuming CP-conservation, the Boltzmann equation here, taking into account all reactions allowed by the $Z_7$ symmetry, is given as:
\begin{equation}
\frac{1}{a^3} \frac{d }{dt}(n_\chi a^3) =  
- \dfrac{3}{n^3_{\chi,\text{eq}}} 
\left( \langle \sigma_{ \chi^* \chi^* \to 5\chi } v \rangle 
+ \langle \sigma_{ \chi \chi \to 5\chi^* } v \rangle\right)
\bigg[ n_\chi^5 - n_\chi^2 n^3_{\chi,\text{eq}}  \bigg] \;.
\end{equation}

\end{enumerate}

We show the resulting unitarity upper limits on the DM mass in the above scenarios in a radiation dominated Universe, in Table~\ref{tab:2}, by requiring that the DM particle saturates the observed DM abundance of $\Omega_{\rm DM} h^2 \simeq 0.12$~\cite{Aghanim:2018eyx}. The upper bounds are shown for three cases in each scenario, with the dominant annihilation mode being from (1) s-wave, (2) p-wave and (3) both s- and p-wave initial states. As the rate for each higher partial wave is expected to be further suppressed by powers of a non-relativistic relative velocity~\cite{Griest:1989wd}, we restrict our considerations to the s- and p-wave contributions only. As we can see from this Table, the well-known upper bound of around $130$ TeV for non-identical s-wave DM pair annihilations to SM states is reproduced. The upper bound for the $3 \rightarrow 2$ scenario is around $1$ GeV, while for the $4 \rightarrow 2$ and $5 \rightarrow 2$ scenarios it is around $7$ MeV and $110$ keV, respectively. For all the $k \rightarrow 2$ scenarios with $k \geq 3$, the bounds do not change significantly on inclusion of higher partial wave contributions to the annihilation rate, since the maximum value of the mass scales as $(2\ell+1)^{1/k}$. For higher values of $k$, the annihilation rates are flux factor suppressed, thereby reducing the unitarity upper limit on the annihilation rate, thus increasing the resulting DM number density, and lowering the upper bound on the DM mass that saturates the observed DM abundance. For non-zero DM spin, angular momentum conservation will impose further selection rules,  allowing only certain annihilation topologies, and within each topology possibly requiring specific values of $\ell$.

 \begin{table}[htb!]
\begin{center}
\def\arraystretch{1.5}
\begin{tabular}{|c|c|c|c|c|}
\hline
 Symmetry & Annihilation channels       & $\ell=0$                 & $\ell=1$                & $\ell=0+1$  \\ \hline
$Z_2$& $\chi + \chi^* \to \text{SM} +\text{SM}$ &   127.7 TeV    &    220 TeV         & 253.5 TeV           \\ \hline
$Z_3$& $3 \chi^{(*)} \to 2  \chi^{(*)}$  &       1.15 GeV    &    1.72 GeV        &  1.91 GeV  \\ \hline
$Z_2$& $4 \chi \to 2 \chi$  &   6.9 MeV           &   9.4 MeV     &   10.1 MeV             \\ \hline
$Z_7$& $5 \chi^{(^*)} \to 2 \chi^{(*)} $  &  114.1 keV     &   139.7 keV     &    147.4  keV           \\ \hline
\end{tabular}
\end{center}
\caption{\label{tab:2} Unitarity upper limits on thermal DM mass in a radiation dominated Universe, for different dominant annihilation topologies of $k \rightarrow 2$ type, with $k \geq 2$. In order to obtain a precise prediction, in each case a minimal scenario with (real or complex) scalar is considered, with a minimal $Z_N$ DM stabilization symmetry. The upper bounds are shown for three cases in each scenario, with the dominant annihilation mode being from (1) s-wave, (2) p-wave and (3) both s- and p-wave initial states.}
\end{table}

We note in passing that there exist constraints on the total energy density of a relativistic species at BBN temperatures (of order MeV) primarily through its effect on the Hubble expansion rate, and also through a small effect due to the kinetic equilibrium of the DM sector with the SM bath, which distributes the heat generated through the $k \rightarrow 2$ annihilation process. 
A DM particle in kinetic equilibrium with the SM sector of mass $\mathcal{O} (\rm MeV)$ or lower will thus be constrained by the considerations of BBN. Evading such constraints will  require additional model-dependent effects affecting the BBN processes.
Since our paper focusses on the production of such DM particles and the model-independent implications of S-matrix unitarity, we did not discuss searches of, or constraints on such DM, as that would depend on the details of the model under consideration. 

\section{Unitarity limits on thermal DM: intermediate matter dominated Universe}
\label{sec_limits_IMD}
For the analyses in the previous section, we have assumed the Universe to be radiation dominated (RD) during the dark-matter freeze-out epoch, and subsequently until the onset of matter-radiation equality at a temperature of around $1$ eV. The successful predictions of big-bang nucleosynthesis (BBN) require the Universe to be RD at temperatures of the order of $1$ MeV~\cite{Gorbunov:2011zz}. However, in the pre-BBN era, the energy density could be dominated by a non-relativistic matter field, which eventually decays to radiation sufficiently before the BBN, restoring back an RD Universe. This requires the heavy matter field ($\Phi$) to be very feebly interacting with the SM bath, such that it is long-lived and does not thermalize with the SM sector, during the epoch of interest. For this section, we further assume that the $\Phi$ field does not possess sufficiently strong number-changing interactions within its own sector either, and explore the consequences of relaxing this assumption elsewhere. Such fields may, for example, interact with the SM sector only gravitationally, as often encountered in extensions of the SM of particle physics. 
\begin{figure} [htb!]
\begin{center} 
\includegraphics[height=5cm]{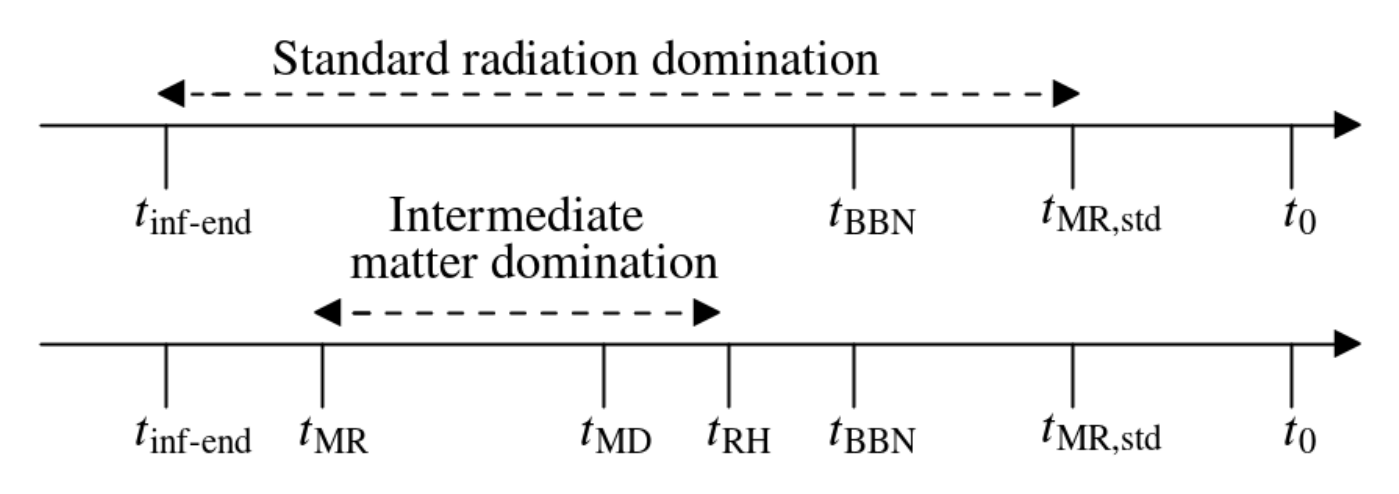}
\caption{\small{\it{Schematic description of the cosmological timeline between the end of inflation $t_{\text{inf-end}}$ and the present epoch $t_0$, in standard cosmology with radiation domination before $t_{\text{MR,std}}$ (top panel), and a modified thermal history with an intermediate period of matter domination (bottom panel), with a nearly constant entropy phase between $t_{\text{MR}}$ and $t_{\text{MD}}$, and varying entropy phase until $t_{\text{RH}}$, when there is reheating back to RD. See text for details.} }}
\label{fig:1}
\end{center}
\end{figure}

If a sufficient density of $\Phi$ is generated at the end of the inflationary reheating along with radiation, then eventually due to the faster dilution of the radiation energy density from the cosmic expansion, compared to that of non-relativistic matter (namely, $\rho_R \propto 1/a^4$ and $\rho_\Phi \propto 1/a^3$ in the absence of any net creation or destruction), the $\Phi$ field energy becomes equal to that of radiation at a time $t=t_{\rm MR}$, say. This results in the onset of a stable matter dominated era which continues during $t_{\rm MR} \lesssim t \lesssim t_{\rm MD}$, when at around $t \simeq t_{\rm MD}$ the decays of the $\Phi$ particles start to produce significant amount of radiation, and thus entropy. This leads to a Universe with energy density dominated by a decaying matter field, during  $t_{\rm MD} \lesssim t \lesssim t_{\rm RH}$, when at around $t \simeq t_{\rm RH}$ most of the $\Phi$ density decays back to radiation, thus restoring the RD phase before the BBN. We schematically show this timeline of cosmological events in Fig.~\ref{fig:1}, in which this alternative thermal history with an intermediate matter dominated (IMD) phase is contrasted with the standard thermal history, assuming an inflationary scenario at the earliest epoch of the Universe. 

The scenario described above is governed by a set of three coupled differential equations, namely, the Boltzmann equations for the energy densities of radiation and $\Phi$, 
\begin{eqnarray}
\frac{d\rho_R (t)}{dt} + 4 H(t) \rho_R (t) &=& \Gamma_\Phi \rho_\Phi (t) \label{Eq_rho_R}\\
\frac{d\rho_\Phi (t)}{dt} + 3 H(t) \rho_\Phi (t) &=& -\Gamma_\Phi \rho_\Phi (t) \label{Eq_rho_Phi},
\end{eqnarray}
and the Friedmann equation determining the Hubble parameter
\begin{equation}
H^2(t) = \frac{8\pi G}{3} \left(\rho_R(t) + \rho_\Phi(t) \right),
\label{Eq_Hubble}
\end{equation}
where, $\Gamma_\Phi$ is the total decay width of $\Phi$, which is assumed to decay entirely to radiation. This system of equations have been studied extensively in the literature within different contexts, including that of inflationary reheating, see for example, Refs.~\cite{McDonald:1989jd, Giudice:2000ex, Gelmini:2006pw, Kolb:1990vq, Gorbunov:2011zz}. We shall adopt a simple analytical approximation, which captures the essence of the modified cosmic history involving the interplay of radiation and $\Phi$ densities, and then treat the evolution of the dark matter density accurately with this cosmology as an input. Our approach largely follows Ref.~\cite{Gorbunov:2011zz}, with certain improvements in the analytical treatment as will be clear from the discussion below.

Eq.~\ref{Eq_rho_Phi} is easily solved to obtain the time-evolution of $\rho_\Phi (t)$ as
\begin{equation}
\rho_\Phi (t) = \rho_\Phi (t_I) \frac{a(t_I)^3}{a(t)^3} e^{-\Gamma_\Phi \left(t-t_I\right)},
\label{Eq_Phi_Energy_Time}
\end{equation}
where $t_I$ is some initial time after inflationary reheating, when the $\Phi$ energy density is $\rho_\Phi (t_I)$. It is convenient to choose $t_I$ in the range $t_{\rm MR} \lesssim t_I \lesssim t_{\rm MD}$, such that during this time the Hubble parameter $H(t)$ is approximately determined by $\rho_\Phi$, with $\Phi$ being an essentially stable non-relativistic matter field. Thus, using Eq.~\ref{Eq_Hubble}, with $H(t) \simeq 2/(3t)$, we find $\rho_\Phi (t_I) \simeq 3 M_P^2 H^2(t_I) \simeq 4M_P^2/(3t_I^2)$, where we have defined the reduced Planck mass by $8\pi G = 1/M_P^2$. With $a(t) \propto t^{2/3}$ during this epoch, we can write using Eq.~\ref{Eq_Phi_Energy_Time}
\begin{equation}
\rho_\Phi (t) = \frac{4M_P^2}{3t^2} e^{-\Gamma_\Phi \left(t-t_I\right)},~~{\rm for} ~t \gtrsim t_{\rm MR}.
\label{Eq_Phi_Energy_Time_2}
\end{equation}

With Eq.~\ref{Eq_Phi_Energy_Time_2} as an input to the Boltzmann equation~\ref{Eq_rho_R} for the radiation energy density, with the boundary condition $\rho_R(t_{\rm MR}) = \rho_\Phi(t_{\rm MR})$, we obtain the approximate solution for $\rho_R(t)$ 
\begin{equation}
\rho_R(t) \simeq \frac{4M_P^2 \Gamma_\Phi}{3} e^{\Gamma_\Phi t_I} \sum_{n=0}^{\infty} \frac{\left(-\Gamma_\Phi \right)^n t^{n-1}}{n !\left(n+\frac{5}{3}\right)} + \frac{4M_P^2 }{3} \frac{t_{\rm MR}^{2/3}}{t^{8/3}},
\label{Eq_Rad_Energy}
\end{equation}
where, the first term in the RHS stems from the decay of the $\Phi$ field, while the second term is the contribution of the radiation density present before $\Phi$ decay. For $t_I \sim \mathcal{O}(t_{\rm MR})$, $\Gamma_\Phi t_I < < 1$, and hence  $e^{\Gamma_\Phi t_I} \sim 1$.

The decay of the $\Phi$ field to radiation generates entropy, which we shall now compute. We define $t_{\rm MD}$ to be the time when
\begin{equation}
\rho_R (t_{\rm MD}) = 2 \times  \frac{4M_P^2 }{3} \frac{t_{\rm MR}^{2/3}}{t_{\rm MD}^{8/3}},
\label{Eq_rad_MD}
\end{equation}
i.e., at $t=t_{\rm MD}$, the contribution to the radiation energy density from $\Phi$ decay equals the pre-existing radiation energy density in Eq.~\ref{Eq_Rad_Energy}. The time scale $ t_{\rm MD}$ is approximately given in terms of $\Gamma_\Phi$ and 
$t_{\rm MR}$ as 
\begin{equation}
t_{\rm MD} \simeq \left(\frac{5}{3 \Gamma_\Phi}\right)^{3/5} t_{\rm MR}^{2/5}.
\end{equation}
Most of the entropy is essentially generated by the time $t=t_{\rm RH}$, which is defined by $\Gamma_\Phi t_{\rm RH}=1$, which is also the time scale for which $\Gamma_\Phi \sim H(t)$. At this time, we find
\begin{equation}
\rho_R(t_{\rm RH}) \simeq 0.332 \times \frac{4M_P^2 \Gamma_\Phi^2}{3} ~~ {\rm with} ~\Gamma_\Phi t_{\rm RH}=1.
\label{Eq_rad_RH}
\end{equation}
Since the entropy density is dominated by relativistic species, we can express the ratio of the total entropy $S(t)$ in a co-moving volume $a(t)^3$ between an initial time $t_i$ and a final time $t_f$ as
\begin{equation}
\frac{S(t_f)}{S(t_i)} = \left(\frac{\rho_R(t_f)}{\rho_R(t_i)}\right)^{3/4} \left(\frac{g_{*,s}(t_f)}{g_{*,s}(t_i)}\right)^{1/4} \left(\frac{a(t_f}{a(t_i)}\right)^3.
\label{Eq_entropy_1}
\end{equation}
Using Eqs.~\ref{Eq_rad_MD} and ~\ref{Eq_rad_RH} in Eq.~\ref{Eq_entropy_1}, with $t_i=t_{\rm MD}$ and $t_f = t_{\rm RH}\sim t_{\rm MD} +\delta$, with $\delta \rightarrow 0$, we obtain the ratio of total entropies in a co-moving volume as
\begin{equation}
\frac{S( t_{\rm RH})}{S(t_{\rm MD})} \simeq \frac{0.26}{\sqrt{\Gamma_\Phi t_{\rm MR}}}.
\label{Eq_entropy_2}
\end{equation}
In the instantaneous decay approximation used above, the relativistic degrees of freedom and the scale factors do not change significantly and therefore cancel out in the ratio in Eq.~\ref{Eq_entropy_1}. 

We can express the product $\Gamma_\Phi t_{\rm MR}$ in Eq.~\ref{Eq_entropy_2} in terms of the temperatures for matter-radiation equality, $T_{\rm MR}$, and for reheating back to RD, $T_{\rm RH}$. For this we use the relations $H(t_{\rm MR})=2\sqrt{2}(2-\sqrt{2})/(3t_{\rm MR})= \sqrt{\rho_R+\rho_\Phi}/(\sqrt{3}M_P)$, and $\Gamma_\Phi  \simeq H(t_{\rm RH})$, to obtain
\begin{equation}
\frac{S( t_{\rm RH})}{S(t_{\rm MD})} \simeq 0.416 \times \frac{T_{\rm MR}}{T_{\rm RH}}.
\label{Eq_entropy_3}
\end{equation}

\subsection{DM freeze-out in constant entropy phase}
If the entropy production happens mostly after the DM freeze-out, i.e., if the DM particle freezes out during the stable matter dominated epoch, or the preceding radiation dominated epoch, then we can express the ratio of the present dark matter number density ($n(T_0)$) and entropy density ($s(T_0)$) in terms of the corresponding ratio at freeze-out, and the entropy dilution factor as follows
\begin{align}
\frac{n(T_0)}{s(T_0)} & = \frac{n (T_{\rm FO})}{s(T_{\rm FO})} \times \frac{S(t_{\rm MD})}{S(t_{\rm RH})} \nonumber \\
& \simeq 2.4 \times \frac{n (T_{\rm FO})}{s(T_{\rm FO})} \times \frac{T_{\rm RH}}{T_{\rm MR}}.
\label{Eq_DM_density_dilution_1}
\end{align}
Thus for these two scenarios of DM freeze-out, we can solve  the Boltzmann equation determining the DM number density in an iso-entropic Universe Eq.~\ref{Eq_Y_isoentropic}, and then obtain the present density using Eq.~\ref{Eq_DM_density_dilution_1}, with the input Hubble parameter for this epoch
\begin{equation}
H_\Phi(T) = \sqrt{\frac{\pi^2 g_*(T)}{90}} T_{\rm MR}^{1/2} \frac{T^{3/2}}{M_P}.
\label{Eq_Hubble_2}
\end{equation}
Since $T_{\rm MR}>T$ in this case, the Hubble expansion is faster than in the corresponding radiation dominated Universe at the same temperature, resulting in an earlier DM freeze-out. The Boltzmann Eq.~\ref{Eq_Y_RD} is then modified for the adiabatic matter dominated Universe as
\begin{equation}
\frac{dY}{dx} =  - \sum_{\rm Channels} (\Delta n_\chi) \frac{\lambda_\Phi}{x^{(3k-2)/2}} \left[Y^k - Y^2 Y_{\rm eq}^{k-2} \right],
\label{Eq_Y_Phi}
\end{equation}
with
\begin{equation}
\lambda_\Phi = \lambda_R \sqrt{\frac{m_\chi}{T_{\rm MR}}}.
\label{Eq_lambda_Phi}
\end{equation}
This equation can be approximately solved for large values of $x$, which then including the entropy dilution factor gives
\begin{equation}
Y_\Phi(x \rightarrow \infty) \simeq \left(\frac{3k-4}{\lambda_\Phi(2k-2)}\right)^{\frac{1}{k-1}} x_F^{\frac{3k-4}{2k-2}} \times \frac{S_i}{S_f}.
\end{equation}

\subsection{DM freeze-out in varying entropy phase}
If the DM freezes out during the phase in which the decay of $\Phi$ is generating significant entropy, within the interval $t_{\rm MD} \lesssim t \lesssim t_{\rm RH}$, the temperature dependence of the Hubble expansion rate is modified to
\begin{equation}
H_{\rm Decay}(T) = \sqrt{\frac{\pi^2 g_*(T_{\rm RH})}{90}}  \frac{T^4}{T_{\rm RH}^{2} M_P},
\label{Eq_Hubble_3}
\end{equation}
where, we have matched the Hubble rate to the one in a radiation-dominated Universe at the boundary $T=T_{\rm RH}$. Further matching the Hubble rate $H_{\Phi}(T)$ and $H_{\rm Decay}(T)$ at $T=T_{\rm MD}$, we find a relation between $T_{\rm MD}$, $T_{\rm MR}$ and $T_{\rm RH}$. Assuming $g_*(T_{\rm MD}) \sim g_*(T_{\rm RH})$, this relation is given by
\begin{equation}
T_{\rm MD} \simeq \left(T_{\rm RH}^2  \sqrt{T_{\rm MR}} \right)^{2/5}.
\label{Eq_TMD}
\end{equation}
 Furthermore, in this scenario, the DM freeze-out is no longer taking place in an iso-entropic Universe. In this case, we can estimate the present DM abundance in terms of the freeze-out abundance as follows. For $T<T_{\rm FO}$, conservation of DM number in a co-moving volume implies that 
\begin{equation}
\frac{n(T_0)}{s(T_0)}  = \frac{n (T_{\rm FO})}{s(T_{\rm FO})} \times \frac{s(T_{\rm FO})}{s(T_{\rm RH})}  \times \frac{a^3(T_{\rm FO})}{a^3(T_{\rm RH})}.
\label{Eq_relic_3}
\end{equation}
For a temperature range in which the variation of the relativistic degrees of freedom in the thermal bath is not significant, we can show that in the decaying matter dominated phase the scale factor $a(T) \propto T^{-8/3}$~\cite{McDonald:1989jd, Gelmini:2006pw}. With this, and using $s(T) \propto T^3$, Eq.~\ref{Eq_relic_3} implies
\begin{equation}
\frac{n(T_0)}{s(T_0)}  \simeq \frac{n (T_{\rm FO})}{s(T_{\rm FO})} \times \frac{T_{\rm RH}^5}{T_{\rm FO}^5}
\label{Eq_dilution_3}
\end{equation}
The Boltzmann equation~\ref{Eq_Boltzmann_2tok} can now be solved with the modified Hubble rate in Eq.~\ref{Eq_Hubble_3} to obtain the number density at freeze-out, which can then be used in Eq.~\ref{Eq_dilution_3} for computing the present DM abundance. 

\subsection{Unitarity constraints in the IMD scenario}
With the above results, we now explore the implications of unitarity on the scenarios of DM freeze-out in which the IMD epoch plays a significant role. There are essentially three free parameters determining DM properties in a thermal history with an IMD phase, $m_\chi$, $T_{\rm MR}$ and $T_{\rm RH}$, where the DM annihilation rate is given entirely in terms of $m_\chi$ by unitarity. For each value of $T_{\rm MR}$ and $T_{\rm RH}$, unitarity thus implies a value of $m_\chi$ that satisfies the observed relic abundance, which is also the highest value of $m_\chi$ allowed in that scenario. The highest value of $m_\chi$ also depends upon whether the DM freeze-out takes place in a nearly constant entropy phase, or in the phase with significant entropy generation. For a given set of $T_{\rm MR}$ and $T_{\rm RH}$ values, this is determined by $m_\chi$, since the temperature at the approximate boundary between these two phases, $T_{\rm MD}$, is obtained through Eq.~\ref{Eq_TMD}.

In order to understand the range of DM mass and annihilation rates allowed by both the relic abundance requirement and unitarity, we parametrize the $2 \rightarrow 2$ reaction rate as 
\begin{equation}
\langle \sigma_{2 \rightarrow 2} v_{\rm rel} \rangle= \langle \sigma_{2 \rightarrow 2} v_{\rm rel} \rangle_0  \sqrt{x}.
\label{Eq_sigma_float_2to2}
\end{equation} 
where, $ \langle \sigma_{2 \rightarrow 2} v_{\rm rel} \rangle_0$ is a free parameter, which takes a maximum value allowed by $s-$wave unitarity as given in Eq.~\ref{Eq_sigma_max_2to2}. Similarly, we parametrize the $3 \rightarrow 2$ reaction rate as 
\begin{equation}
\langle \sigma_{3 \rightarrow 2} v_{\rm rel}^{2} \rangle= \langle \sigma_{3 \rightarrow 2} v_{\rm rel}^{2} \rangle_0   x^{2},
\label{Eq_3to2_sigma_float}
\end{equation}
where, $ \langle \sigma_{3 \rightarrow 2} v_{\rm rel}^{2} \rangle_0$ is a free parameter, which takes a maximum value allowed by $s-$wave unitarity as given in Eq.~\ref{Eq_3to2_sigma_max}. Finally, the $4\rightarrow 2$ reaction rate is parametrized by 
\begin{equation}
\langle \sigma_{4 \rightarrow 2} v_{\rm rel}^{3} \rangle= \langle \sigma_{4 \rightarrow 2} v_{\rm rel}^{3} \rangle_0  x^{7/2},
\label{Eq_4to2_sigma_float}
\end{equation}
where again $ \langle \sigma_{4 \rightarrow 2} v_{\rm rel}^{3} \rangle_0$ is a free parameter, whose maximum value allowed by $s-$wave  unitarity is as given in Eq.~\ref{Eq_4to2_sigma_max}.

\begin{figure} [htb!]
\begin{center} 
\includegraphics[scale=0.55]{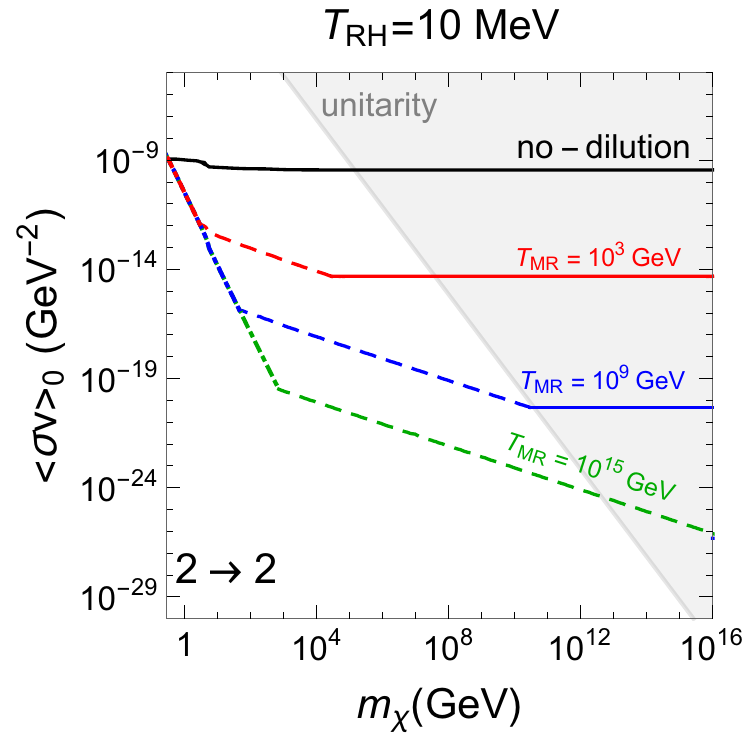} \hspace{0.5cm}
\includegraphics[scale=0.55]{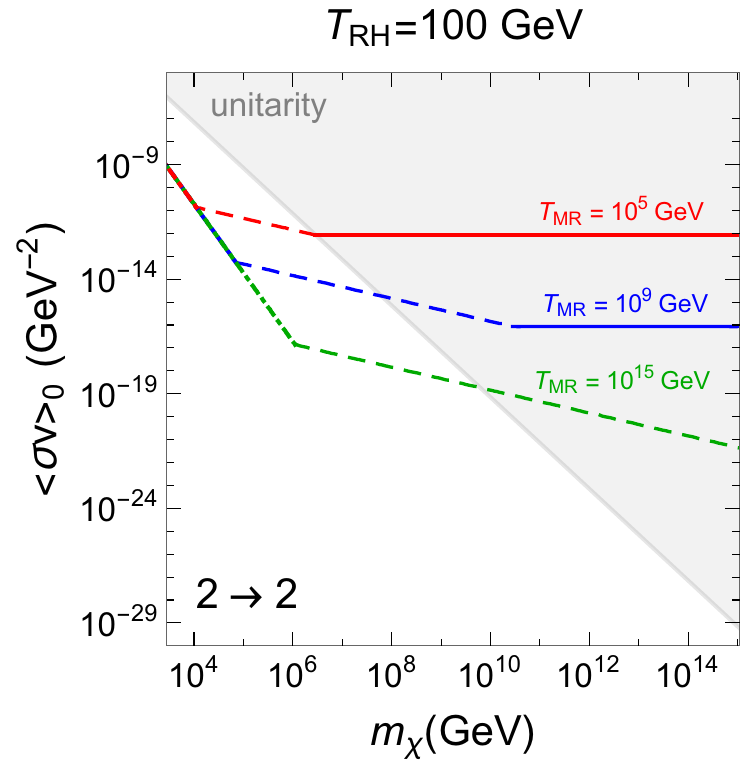}
\end{center}
\caption{\small{\it{Region in the $ \langle \sigma_{2 \rightarrow 2} v_{\rm rel} \rangle_0$ and $m_\chi$ plane in which the observed value of the DM density $\Omega_{\rm DM} h^2=0.12$ is satisfied, for dominantly $2 \rightarrow 2$ DM annihilations, with an intermediate matter dominated epoch, for the scenarios with $T_{\rm RH}=10$ MeV (left panel) and with  $T_{\rm RH}=100$ GeV (right panel). The results are shown for different values of the matter-radiation equality temperature $T_{\rm MR}$. In each case, DM freeze-out may take place during radiation domination (solid lines), nearly stable matter domination (dashed lines), or decaying matter domination epochs (dot-dashed lines). The results in a purely radiation dominated scenario, with no subsequent IMD or entropy dilution are also shown for comparison (black solid line).} } }
\label{Fig_2to2_IMD}
\end{figure}
We show in Fig.~\ref{Fig_2to2_IMD} the region in the $ \langle \sigma_{2 \rightarrow 2} v_{\rm rel} \rangle_0$ and $m_\chi$ plane in which the observed value of the DM density $\Omega_{\rm DM} h^2=0.12$ is reproduced. The scenario with $T_{\rm RH}=10$ MeV is shown in the left panel, while the one with $T_{\rm RH}=100$ GeV is shown in the right panel. In each figure we show the results for different values of the matter-radiation equality temperature $T_{\rm MR}=10^3, 10^9$ and $10^{15}$ GeV using the red, blue and green lines, respectively, for the $T_{\rm RH}=10$ MeV scenario. To generate a minimum amount of entropy dilution that can impact the DM properties, the corresponding $T_{\rm MR}$ values chosen for $T_{\rm RH}=100$ GeV starts from $T_{\rm MR}=10^5$ GeV (red line in the right panel), the other two  $T_{\rm MR}$ values shown being the same. For comparison, the black solid line (left panel) shows the results in a purely radiation dominated scenario, with no subsequent IMD or entropy dilution. The grey shaded region is disallowed by the unitarity limit on the annihilation rate in Eq.~\ref{Eq_sigma_max_2to2}. As we can see from this figure, for a given $T_{\rm MR}$  and $T_{\rm RH}$, the required value of $ \langle \sigma_{2 \rightarrow 2} v_{\rm rel} \rangle_0$ changes with different slopes for increasing values of $m_\chi$, depending upon whether the DM freeze-out takes place during radiation domination (shown by solid lines), nearly stable matter domination (dashed lines), or decaying matter domination epochs (dot-dashed lines). This is due to the differences in the expansion rate of the Universe in these regions, as shown by the modification to the Hubble rate compared to a RD scenario in Eqs.~\ref{Eq_Hubble_2} and ~\ref{Eq_Hubble_3}. For freeze-out during stable or decaying matter domination, the net DM yield depends on $m_\chi$ explicitly, in addition to the dependence through $x_F$, and higher values of $m_\chi$ require lower cross-sections to saturate the observed density. The maximum cross-section allowed by unitarity also decreases with increasing $m_\chi$. The intersection point of the relic density allowed line and the unitarity constraint thus gives the maximum allowed mass for a given scenario. In the purely RD scenario (black solid line, left panel), this intersection point gives the unitarity limit quoted in Table~\ref{tab:2}. For a fixed $T_{\rm RH}$, increasing $T_{\rm MR}$ leads to a higher dilution of the frozen out DM density. To compensate for that, we need higher DM densities to survive at freeze-out. This implies an earlier decoupling, thereby requiring lower cross-sections for the same mass to satisfy the DM abundance. This is why the unitarity limit on the DM mass that satisfies the relic abundance also increases for higher values of $T_{\rm MR}$. 

Our results in Fig.~\ref{Fig_2to2_IMD} can be compared with the results in Ref.~\cite{Delos:2019dyh}. The required values of the DM mass and annihilation rates that satisfy the observed DM abundance are in agreement in the two studies. However, in Fig.~\ref{Fig_2to2_IMD}, the unitarity constraints are shown accurately, while Ref.~\cite{Delos:2019dyh} presented only an estimate using $\mathcal{O}(1)$ coupling values, and hence the unitarity disallowed regions differ accordingly.

\begin{figure} [htb!]
\begin{center} 
\includegraphics[scale=0.55]{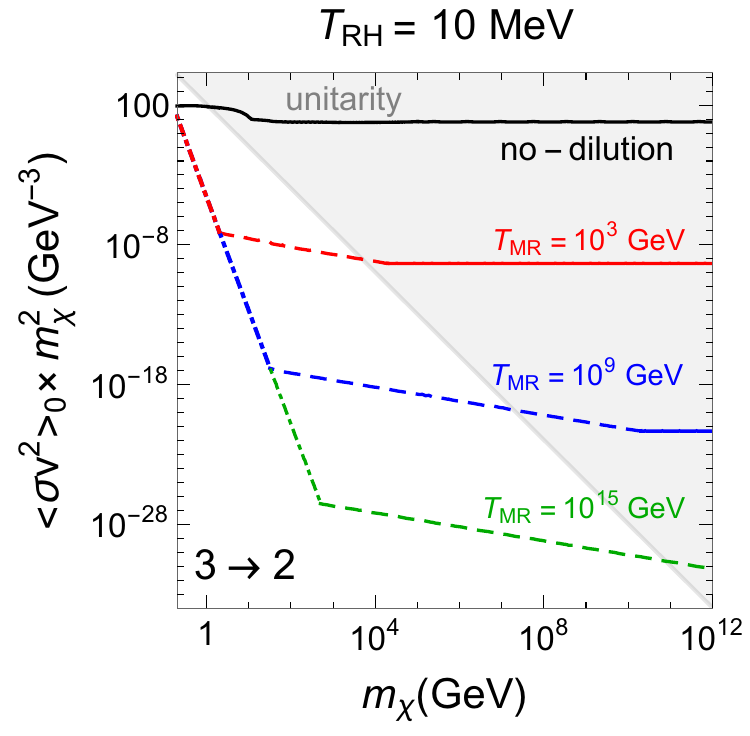} \hspace{0.5cm}
\includegraphics[scale=0.55]{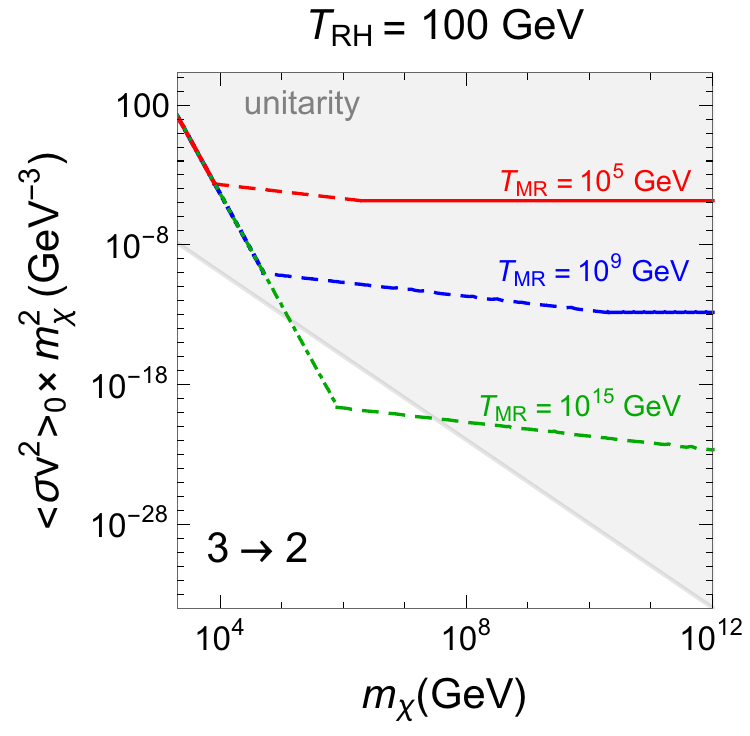}
\end{center}
\caption{\small{\it{Same as Fig.~\ref{Fig_2to2_IMD}, for a scenario in which the dominant number-changing DM annihilations are of $3 \rightarrow 2$ type.} } }
\label{Fig_3to2_IMD}
\end{figure}
\begin{figure} [htb!]
\begin{center} 
\includegraphics[scale=0.55]{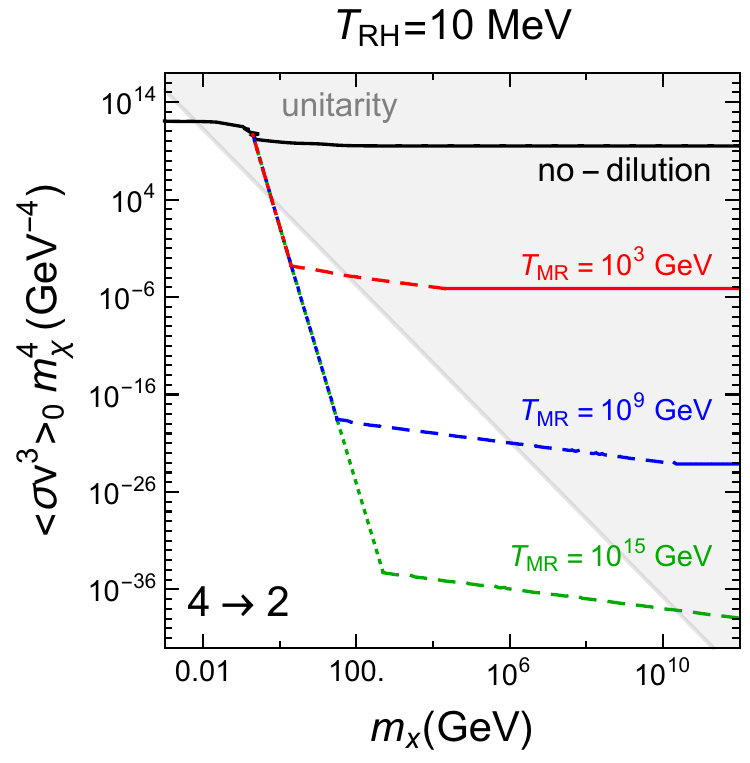} \hspace{0.5cm}
\includegraphics[scale=0.55]{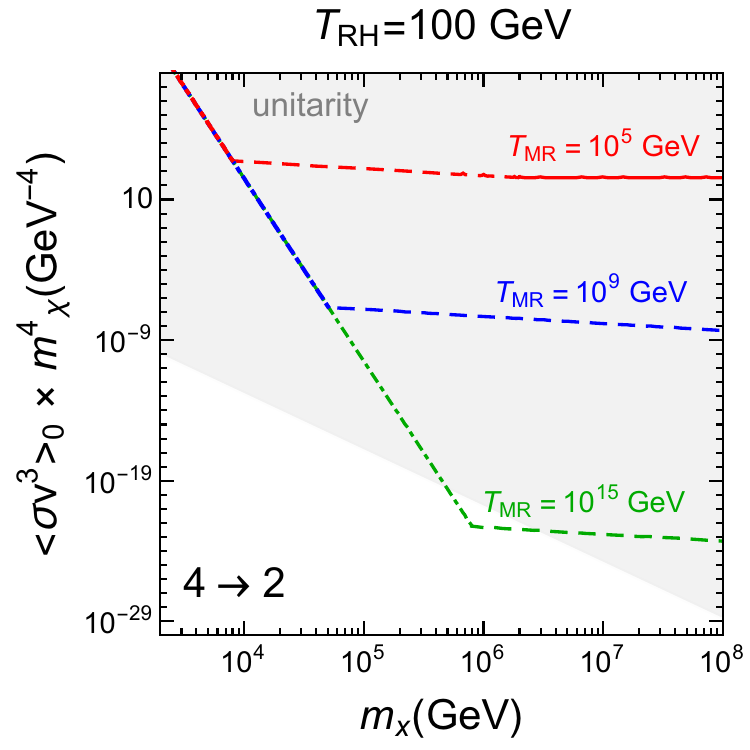}
\end{center}
\caption{\small\it{{Same as Fig.~\ref{Fig_2to2_IMD}, for a scenario in which the dominant number-changing DM annihilations are of $4 \rightarrow 2$ type.} } }
\label{Fig_4to2_IMD}
\end{figure}

The results for the scenarios in which $3 \rightarrow 2$ and $4 \rightarrow 2$ annihilations are the dominant DM number changing process, are qualitatively similar to the $2 \rightarrow 2$ scenario discussed above, as shown in Figs.~\ref{Fig_3to2_IMD} and~\ref{Fig_4to2_IMD}, respectively. Since the mass dependence of the maximum annihilation rates in these scenarios are stronger than in the $2 \rightarrow 2$ case, the implied upper bounds on the DM mass are also correspondingly smaller. 

\begin{figure} [htb!]
\begin{center} 
\includegraphics[scale=0.36]{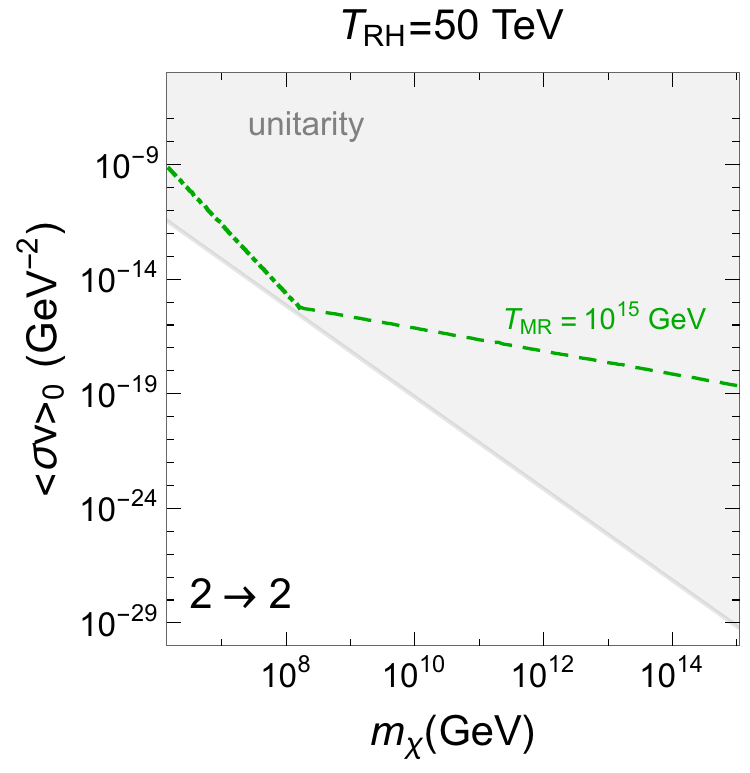} \hspace{0.2cm}
\includegraphics[scale=0.38]{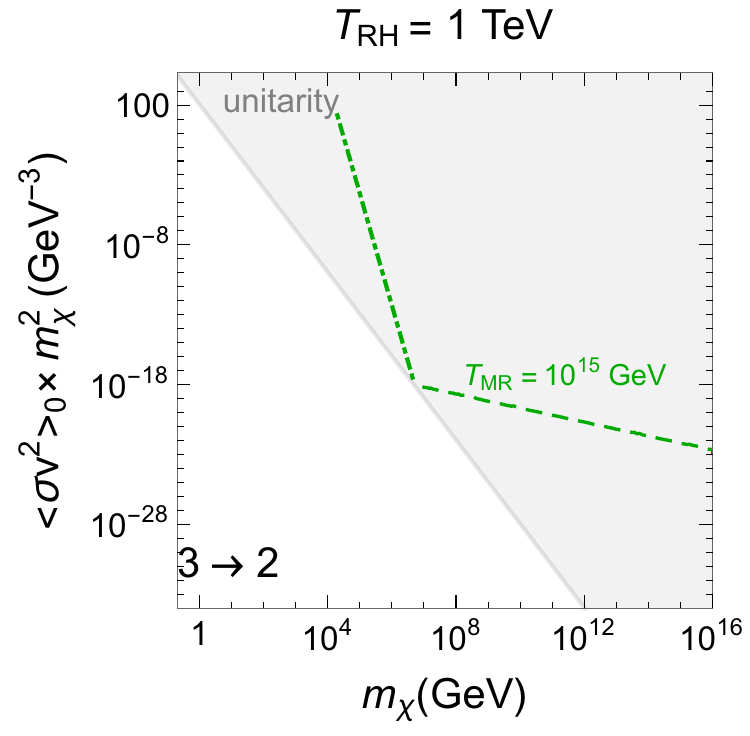} \hspace{0.2cm}
\includegraphics[scale=0.38]{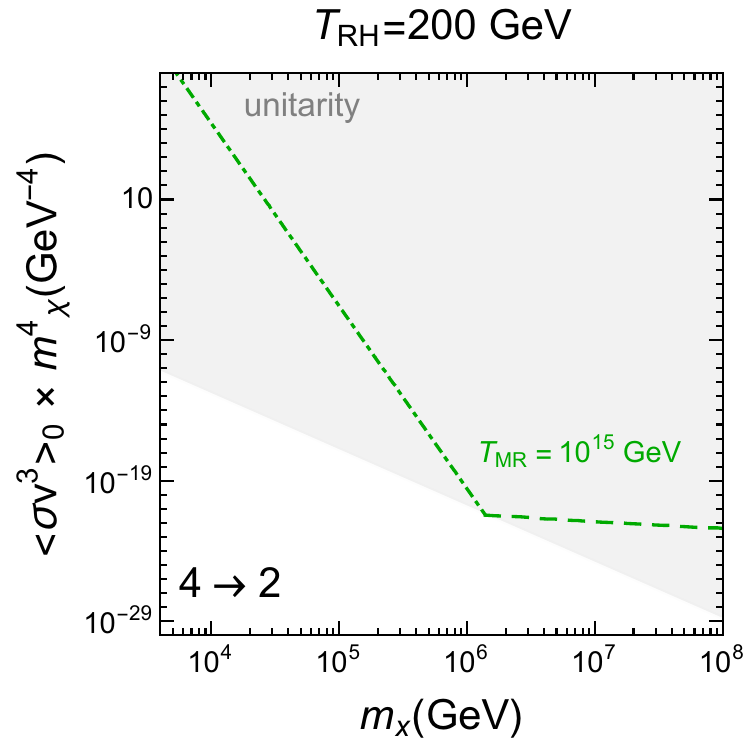} \hspace{0.2cm}
\end{center}
\caption{\small{\it{Approximate maximum value of $T_{\rm RH}$ for which no $T_{\rm MR}$ values upto $10^{15}$ GeV can accommodate the requirements of both DM relic density and unitarity, for the scenarios in which the dominant number-changing DM annihilations are of $2 \rightarrow 2$ (left panel), $3 \rightarrow 2$ (middle panel) and $4 \rightarrow 2$ (right panel) type. For details, see Fig.~\ref{Fig_2to2_IMD} and the related text.} } }
\label{Fig_TRH_Bound}
\end{figure}

In order for the IMD epoch to have any impact on the properties of DM, the freeze-out temperature $T_{\rm FO}$ should be larger than $T_{\rm RH}$. This in turn implies a typical minimum value of the DM mass of around $x_F T_{\rm RH}$, with the value of $x_F$ usually lying in the range of $20-30$. However, the cross-sections required by the relic abundance condition might already be higher than the unitarity upper limit for this mass. Thus we find that with increasing $T_{\rm RH}$, the region allowed by the combination of relic density and unitarity conditions gets reduced. This is more pronounced for the $3 \rightarrow 2$ and $4 \rightarrow 2$ annihilation scenarios, in which, as we can see in Figs.~\ref{Fig_3to2_IMD} and~\ref{Fig_4to2_IMD}, for $T_{\rm RH}=100$ GeV, only a small region remains allowed even with $T_{\rm MR}$ as high as $10^{15}$ GeV. How high can $T_{\rm RH}$ be then for this largest value of $T_{\rm MR}=10^{15}$ GeV considered in this study? We show in Fig.~\ref{Fig_TRH_Bound} an order of magnitude estimate for the highest possible $T_{\rm RH}$ value. As we can see from this figure, while for dominantly $2 \rightarrow 2$ annihilating DM the maximum $T_{\rm RH} = \mathcal{O}(50)$ TeV, for dominantly $3 \rightarrow 2$ and $4 \rightarrow 2$ annihilating DM, one obtains a much stronger upper bound of $T_{\rm RH} = \mathcal{O}(1)$ TeV and $T_{\rm RH} = \mathcal{O}(200)$ GeV, respectively. Thus no scenario with an IMD epoch is viable above these values of $T_{\rm RH}$ if $T_{\rm MR} \lesssim 10^{\rm 15}$ GeV. We have chosen the maximum value of $T_{\rm MR}$ to be at around the grand unification scale. 
Considerations of the maximum value of the Hubble expansion rate in inflationary scenarios, that are consistent with the upper bound on primordial gravitational waves, can also lead to similar restrictions on the scale of $T_{\rm MR}$~\cite{Drees:2017iod}. 

\section{Summary}
\label{sec_summary}
In this paper, we have studied the implications of $S-$matrix unitarity on general $k \rightarrow 2$ thermal DM annihilations, for $k \geq 2$. We first derived the upper limit on the inelastic $2 \rightarrow k$ annihilation cross-sections using the optical theorem, and the matrix element and cross-section for the elastic $2 \rightarrow 2$ annihilation process. These limits are obtained both for non-identical and identical initial state particles, where for the latter case they are larger by a factor of two, with two particles in the initial state. The thermally averaged $k \rightarrow 2$ annihilation rates in the Boltzmann equation are expressed in terms of the $2 \rightarrow k$ annihilation rates using detailed balance. On thermal averaging, $\langle \sigma_{2 \rightarrow k} v_{\rm rel} \rangle$ turns out to be the same  for both the identical and non-identical initial state particles. The unitarity limits on the annihilation rates translate to upper limits on the DM mass, which generalize the well-known result for $2 \rightarrow 2$ DM pair annihilation to a pair of SM particles, with an upper bound of around $130$ TeV for non-identical s-wave DM pair annihilations. The s-wave upper bound for the $3 \rightarrow 2$ scenario is found to be around $1$ GeV, while for the $4 \rightarrow 2$ and $5 \rightarrow 2$ scenarios it is around $7$ MeV and $110$ keV, respectively. For all the $k \rightarrow 2$ scenarios with $k \geq 3$, the bounds do not change significantly on inclusion of higher partial wave contributions to the annihilation rate. 

We also consider the consequences of possible departures from a radiation dominated Universe in the pre-BBN era, with an intermediate period of matter domination, in which the DM freeze-out may take place either when the heavy matter field is mostly stable, or during a time when its decays are producing a significant amount of entropy. The dual effects of modified temperature dependence of the Hubble parameter in these epochs, and the dilution of the frozen-out DM number density due to the generation of entropy, leads to modification of the unitarity limits on the DM mass, which are weaker than in the radiation dominated scenario. These limits depend upon the reheating temperature $T_{\rm RH}$ in which radiation domination is restored, as well as the net entropy dilution factor which, in addition to the reheating temperature also depends on the temperature $T_{\rm MR}$ at the beginning of the IMD epoch. For a given $T_{\rm RH}$, higher values of $T_{\rm MR}$ naturally lead to correspondingly higher limits on the DM mass, due to a larger effective dilution factor. However, on the other hand, increasing $T_{\rm RH}$ leads to a reduction of the region allowed by the combined constraints of the observed DM abundance and unitarity, eventually strongly disfavouring values of reheating temperature higher than $\mathcal{O}(200)$ GeV for $k \geq 4$, $\mathcal{O}(1)$ TeV  for $k = 3$ and $\mathcal{O}(50)$ TeV for $k=2$, with $T_{\rm MR} \lesssim 10^{15}$ GeV.

\section*{Acknowledgements}
SM would like to thank Deep Ghosh for many helpful discussions and important inputs, especially on the derivation of the partial-wave unitarity results in QFT, and Avirup Ghosh and Shigeki Matsumoto for several useful discussions, comments and suggestions. DB would like to thank Roman Walczak for helpful correspondence regarding partial-wave analysis and its implications in non-relativistic physics,  and Cedric Delaunay for useful discussions.

\end{document}